\documentclass[journal, onecolumn]{IEEETran}

\usepackage[utf8]{inputenc} 
\usepackage[T1]{fontenc}    
\usepackage{mdframed}
\usepackage{cite}
\usepackage{amsmath,amssymb,amsfonts,amsthm}
\usepackage{algorithmic}
\usepackage{graphicx}
\usepackage{textcomp}
\usepackage{xcolor}
\usepackage{hyperref} 
\usepackage{multirow} 
\usepackage{tabularx} 
\usepackage{subcaption}
\usepackage{caption}
\usepackage{multirow} 
\usepackage[inline]{enumitem}
\def\BibTeX{{\rm B\kern-.05em{\sc i\kern-.025em b}\kern-.08em
    T\kern-.1667em\lower.7ex\hbox{E}\kern-.125emX}}
\usepackage{colortbl}
\usepackage{xcolor}
\usepackage{tabularx}
\usepackage{booktabs}
\usepackage{titlesec}
\usepackage{tikz}
\usetikzlibrary{positioning, shapes.geometric, arrows.meta, fit}
\usepackage{easyReview}
\usepackage{xcolor}
\usepackage{pgfplots}
\pgfplotsset{compat=1.18} 
\usepackage{siunitx} 
\usepackage{lipsum}
\newmdenv[
    backgroundcolor=gray!10,
]{conclusionbox}
\usepackage{float}
\usepackage{graphicx}
\graphicspath{{media/}}     

\begin{document}
\title{Profiling Concurrent Vision Inference Workloads on NVIDIA Jetson - Extended}
\author{
\IEEEauthorblockN{Abhinaba Chakraborty\IEEEauthorrefmark{1}
                  Wouter Tavernier\IEEEauthorrefmark{1}
                  Akis Kourtis \IEEEauthorrefmark{2}
                  Mario Pickavet\IEEEauthorrefmark{1}
                  Andreas Oikonomakis \IEEEauthorrefmark{2}
                  Didier Colle\IEEEauthorrefmark{1}
                  }\\
\IEEEauthorrefmark{1}ID Lab, University of Ghent-imec, Ghent, Belgium \\
\IEEEauthorrefmark{2}NCSRD, Demokritos, Athens, Greece
}

\maketitle

\begin{abstract}
The proliferation of IoT devices and advancements in network technologies have intensified the demand for real-time data processing at the network edge. To address these demands, low-power AI accelerators, particularly GPUs, are increasingly deployed for inference tasks, enabling efficient computation while mitigating the latency and bandwidth limitations of cloud-based systems. Despite their growing deployment, GPUs remain underutilised even in computationally intensive workloads. This underutilisation stems from the limited understanding of GPU resource sharing, particularly in edge computing scenarios. In this work, we conduct a detailed analysis of both high- and low-level metrics, including GPU utilisation, memory usage, streaming multiprocessor (SM) utilisation, and tensor core usage, to identify bottlenecks and guide hardware-aware optimisations. By integrating traces from multiple profiling tools, we provide a comprehensive view of resource behaviour on NVIDIA Jetson edge devices under concurrent vision inference workloads. Our findings indicate that while GPU utilisation can reach $100\%$ with specific optimisations, critical low-level resources, such as SMs and tensor cores, often operate at only $15\%$ to $30\%$ utilisation. Moreover, we observe that certain CPU-side events, such as thread scheduling and context switching, frequently become bottlenecks, further constraining overall GPU performance. We provide several key observations for users of vision inference workloads on NVIDIA edge devices.
\end{abstract}


\section{Introduction}\label{sec:introduction}
\noindent
 The expanding IoT ecosystem generates a vast amount of data requiring real-time processing for applications like autonomous vehicles \cite{autonomous_vehicle_1, autonomous_vehicles_2} and smart cities \cite{smart_cities, smart_city_2}. Traditional cloud services \cite{google_cloud, aws, cloud_services} often face challenges like latency and privacy, making edge computing with GPUs \cite{understanding_gpu_architecture} a compelling alternative. Despite advances in GPU architectures(such as the introduction of tensor cores \cite{tensorcore}), optimal resource utilisation remains elusive, particularly during complex interactions with deep learning (DL) \cite{hands_on_machine_learning, intro_machine_learning,ian_deeplearning, deep_neural} models.

\noindent
To optimise performance in edge computing inference systems, it is imperative to strike a balance between architectural resource utilisation and workload selection of optimised workloads \cite{resource_utilization_1, resource_utilization_2}. Research in this domain primarily focuses on two directions: developing systems tailored to accommodate workloads on edge devices and designing high-performance, energy-efficient accelerators. For the former, a pivotal decision involves determining whether tasks should be executed locally at the edge or offloaded to the cloud \cite{cloud_offload}. For instance, in cloud environments equipped with NVIDIA A40 GPUs, a single $YoloV8n$ \cite{yolov8n} model can process over $1000$ images per second at $fp16$ precision. However, network-related delays, encompassing both transmission and processing overheads, reduce the effective throughput. Rather than relying on an iterative trial-and-error method to meet quality-of-service (QoS) requirements, decisions can be guided by offline performance analysis. Tasks may be offloaded to the cloud or distributed across additional AI accelerators to achieve effective load balancing.

\begin{figure}[H]
    \centering
    \includegraphics[width=\columnwidth]{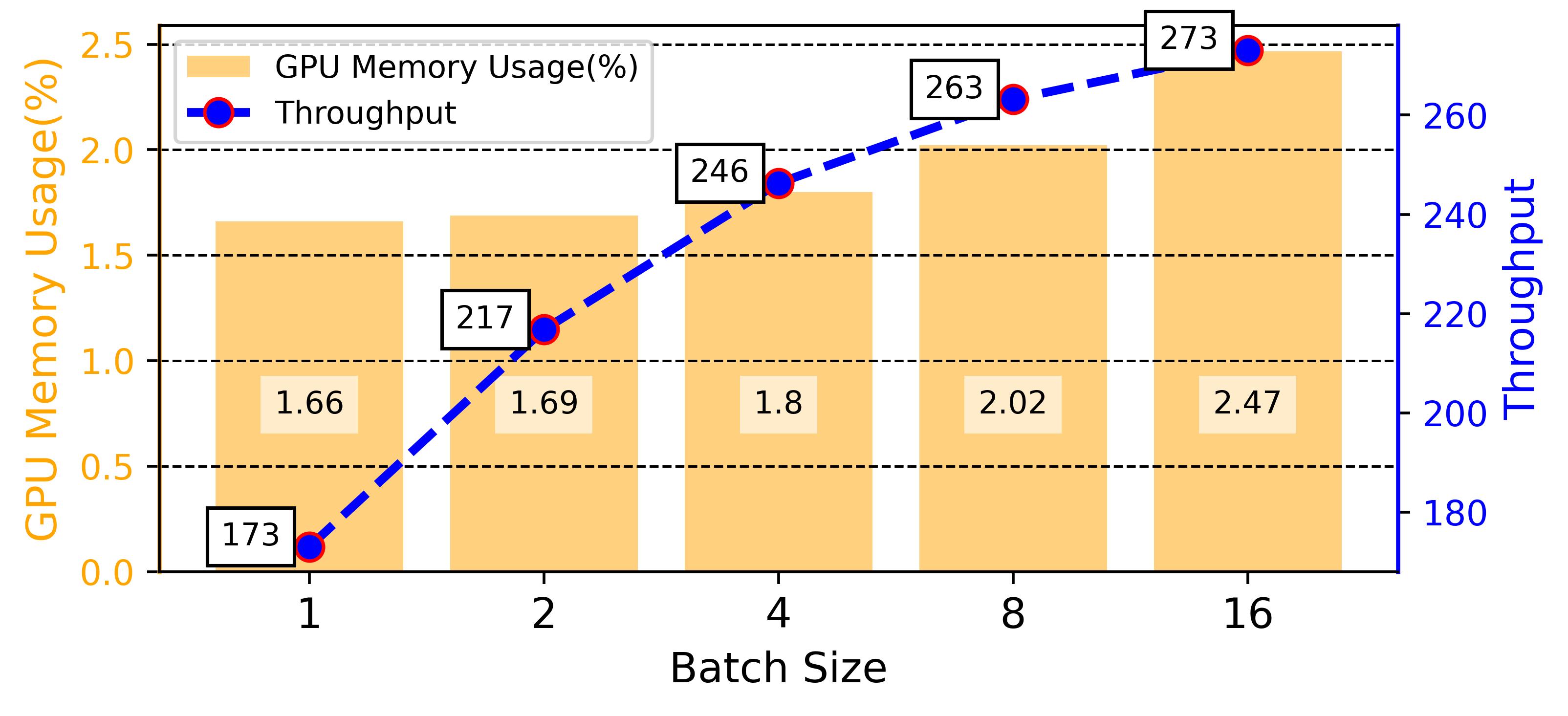}
    \caption{GPU Memory Usage and Throughput vs. Batch Size for \textit{ResNet50} $fp16$ precision model}
    \label{gpu_mempry_throughput_intro_resnet}
\end{figure}

\noindent
DL compilers \cite{deep_learning_compiler} and SDKs like TensorRT \cite{tensorrt} optimise models for specific platforms, balancing accuracy, energy consumption, and other factors that are crucial for edge computing \cite{edge_computing}. These optimisations reduce workload overhead, enabling concurrent model processing on the same edge GPU. However, the limited capacity of edge devices often requires manual trial and error for designing an edge system architecture. Orchestration frameworks like Kubernetes \cite{kubernetes} and YARN \cite{yarn} often prohibit GPU sharing, dedicating one GPU per DL process, which leads to underutilization \cite {9428512}. For example, deploying a \textit{ResNet50} \cite{resnet50} model with \textit{fp16} mixed-precision\footnote{Weight precision refers to the accuracy of representing a DL model's weights, usually in terms of number of bits.} on an NVIDIA Jetson Orin Nano device \cite{jetson} with TensorRT optimisations shows over $98\%$ GPU utilisation while GPU memory usage remains below $3\%$. Throughput\footnote{\textit{Throughput} is the number of images processed per unit time.} increases with batch size\footnote{\textit{Batch\ size} is the number of images processed per inference loop.} but levels off at higher values, as illustrated in Fig-\ref{gpu_mempry_throughput_intro_resnet}.
Power consumption is also a key consideration for edge devices during runtime. These complexities necessitate a thorough study of various metrics and a comparison of different tools, making deployment and DL model design optimisation decisions intricate and labour-intensive.

\noindent
To this end, this paper provides an in-depth analysis of computing resource usage of two NVIDIA Jetson devices (Jetson Orin Nano \cite{jetson}, Jetson Nano \cite{jetson_nano}), focusing on the impact of different weight precisions, batch sizes and multiple concurrent DL processes on metrics such as throughput, energy consumption, and other low-level GPU metrics. Key contributions include:
\begin{enumerate}[]
    \item Analysis of GPU-accelerated vision inference workloads on NVIDIA Jetson devices and identification of relevant metrics using existing profiling tools.
    \item Evaluation of the impact of different numeric precision levels, batch sizes, and numbers of concurrent processes across the SoC, GPU, and kernel levels in modern edge GPUs.
    \item Identification of runtime bottlenecks arising from varying batch sizes and concurrent execution levels, with insights into their implications for system performance and scalability.
\end{enumerate}
In the following sections, we discuss background (Section \ref{related_works}). We then provide an overview of our experimental setup and methodology (Sections \ref{experimental_setup} and \ref{methodology}). In Section \ref{collected_metrics} we provide detailed collected metrics and provide importance in different contexts. Section-\ref{results} presents a detailed workload analysis of our extensive experiments. Section-\ref{kernel_level} provides an overview of kernel-level analysis and potential bottlenecks during runtime execution.
\section{Background}\label{related_works}
\noindent
Deep Neural Networks (DNNs) \cite{deep_neural} are increasingly employed in edge AI applications such as classification \cite{ian_deeplearning}, object detection \cite{object_detection} etc., where we extract features and classify inputs. Traditionally, cloud offloading solutions \cite{edge_offload_issues} rely on high bandwidth and efficient resource allocation. Recently, inference-on-edge \cite{inference_on_edge_1, inference_on_edge_2} has emerged as a promising alternative, offering reductions in latency, bandwidth consumption, and power usage. Techniques such as model redesign and custom architectures are driving advancements in edge-based inference. However, the diversity of hardware and software environments complicates the development of universally optimised frameworks. On general-purpose hardware, linear algebra libraries like $BLAS$ \cite{blas_library} enable efficient deep-learning computations, but manual optimisation for each hardware configuration remains a laborious process. To mitigate this, domain-specific deep learning compilers and frameworks such as $TVM$ \cite{tvm_compiler}, $Glow$ \cite{glow_compiler}, $nGraph$ \cite{ngraph_compiler}, $XLA$ \cite{xla_compiler}, and TensorRT \cite{tensorrt} have gained prominence. These compilers optimise model-specific implementations for diverse hardware, employing techniques like layer and operator fusion to enable efficient code generation.

\noindent
Among edge platforms, GPUs are a prevalent architecture for inference due to their broad support. Existing research stresses the importance of concurrent execution within GPUs to maximise their capabilities \cite{benchmark_analysis_jetson_tx2, nestdnn}. GPUs managing multiple concurrent applications must support some form of virtualisation. Application-level concurrency is a relatively recent development, primarily supported on cloud-based GPUs. The two prevalent approaches are $time\ multiplexing$ and $space\ multiplexing$. Early efforts employed time multiplexing, interleaving applications at predetermined scheduling points \cite{gpuvim}. However, this approach resulted in severe performance degradation as the number of concurrent applications increased \cite{mask}. NVIDIA introduced an alternative approach called $Multi-Process\ Service$ (MPS) \cite{mps_quitting, mps_1}, which employs spatial multiplexing. Under this scheme, different applications are assigned distinct partitions within the same address space, with isolation guaranteed as long as no illegal memory accesses occur. The latest NVIDIA GPU architectures, Turing \cite{turing_gpu} and Ampere \cite{ampere_gpu}, extend this basic MPS support to provide full address space isolation. Unfortunately, Jetson GPUs \cite{jetson, jetson_nano}, which are state-of-the-art for edge applications, do not support MPS. As a result, these devices must rely on either space or time multiplexing. Often, Jetson processors feature integrated unified memory systems wherein RAM is shared between the GPU and CPU. While this type of design eliminates communication overhead between the CPU and GPU, it also leads to rapidly increasing memory consumption as the number of processes scales.

\noindent
Consequently, a deep understanding of system capabilities is imperative when designing inference systems. Tools such as NVIDIA Triton Server \cite{triton-pref} provide a high-level performance overview but fail to capture the utilisation of internal GPU components. Several studies have examined Jetson devices under vision workloads. For instance, \cite{jetson_performance_1} evaluated the Jetson Xavier NX \cite{jetson_xavier} for federated learning applications, while \cite{jetson_performance_2} demonstrated a $16.1\%$ speed-up using optimisations on these edge devices. Similarly, \cite{jetson_performance_3} assessed Jetson platforms based on floating-point operations, and \cite{jetson_performance_4} evaluated a range of algorithms on these platforms.

\noindent
Comprehensive high-level performance analyses of edge platforms, including Jetson, have been provided in \cite{deep-edge-bench, quant_analysis, backboneanalysis}. Meanwhile, the authors of \cite{ai_multi_tenancy} empirically demonstrated that inference throughput could increase by up to $3.8\times$ when running concurrent deep learning applications on edge devices. In \cite{performance_and_diksha} authors provided a micro-architectural characterisation of concurrent executions in GPUs.

\noindent
Research on Jetson devices has primarily focused on high-level analysis using lightweight profiling tools. However, there has been limited exploration of low-level metrics, such as streaming multiprocessor (SM) or tensor core utilisation, particularly in scenarios with concurrent workloads. While high-level insights are undoubtedly important, low-level metrics are crucial for uncovering performance bottlenecks, guiding hardware improvements, and enhancing fault tolerance. In this study, we provide a detailed examination of both high-level and low-level metrics, including GPU utilisation, memory usage, SM activity, and tensor core performance. We also identify key bottlenecks in concurrent application scenarios.


\section{Experimental Setup}\label{experimental_setup}
\subsection{Target Platform}
For our experiments, we use two NVIDIA Edge Jetson GPUs: the \textit{NVIDIA Jetson Orin Nano} \cite{jetson} and the \textit{Jetson Nano} \cite{jetson_nano}, based on the Ampere \cite{ampere_gpu} and Maxwell \cite{nvidia_maxwell} architectures, respectively. The important specifications of the target platforms are shown in the Table~\ref{spec_table}.
\begin{table}[b!]
\centering
\begin{tabularx}{\columnwidth}{>{\raggedright\arraybackslash}p{1in} >{\raggedright\arraybackslash}X >{\raggedright\arraybackslash}X} \midrule
\multicolumn{3}{c}{\textbf{Edge GPU Specification}} \\
\rowcolor{gray!20} \textbf{Metric} & \textbf{Jetson Orin Nano} & \textbf{Jetson Nano} \\
CPU & 6-core Arm Cortex-A78AE \cite{arm_ref_manual} & 4-core ARM Cortex-A57 \cite{arm_a57} \\
GPU & 1024-core Ampere & 128 core Maxwell\\
Tensor Cores & 32 & - \\ 
Unified Memory & 8GB & 4GB \\
Power & 7-15W & 5-10W \\
\bottomrule
\end{tabularx}
\caption{NVIDIA Jetson GPUs}
\label{spec_table}
\end{table}

The Jetson Nano is the entry-level option in NVIDIA's Jetson range, offering the lowest performance. In contrast, the Jetson Orin Nano incorporates cutting-edge GPU architecture with Tensor Cores, offering advanced computational capabilities. While other models in the Orin series are available, their differences primarily lie in performance, with variations in GPU frequency and the number of Tensor Cores. We believe that performance insights and bottlenecks observed in the Jetson Orin Nano can be extrapolated to the broader Orin series. Together, these two devices provide compelling results, offering valuable insights into performance enhancements and limitations within the edge computing paradigm, particularly for computer vision inference workloads.
\subsection{Tools and Libraries}
To emulate a multi-process environment with varying batch sizes, we utilise \textit{trtexec} \cite{trtexec}, a command-line tool within the TensorRT (TRT) SDK. This tool enables the generation and execution of TRT models under various conditions while providing high-level memory and timing data for inference workloads. For detailed application profiling, including kernel and CUDA API \cite{cuda_manual} performance, we use \textit{Nsight Systems} \cite{nisght_systems}. Additionally, the \textit{Jetson-Stats} library \cite{jetson_stats} is used to monitor GPU usage and power consumption, providing real-time insights.

\subsection{Vision Workloads}
In our work, we focus on vision workloads using three distinct deep-learning models for image classification, segmentation, and object detection. For classification and segmentation, we select the \textit{ResNet50} \cite{resnet50} and \textit{FCN\_ResNet50} \cite{7298965} models, respectively, obtaining their implementations and weights from the PyTorch \cite{pytorch_material} Hub. For object detection, we use the \textit{YoloV8n} \cite{yolov8n} model from Ultralytics \cite{ultralytics}, also based on PyTorch. We convert these PyTorch models to ONNX \cite{onnxruntime}, then to TensorRT (TRT) models. During TRT model generation, we optimise for specific batch sizes by disabling dynamic batching and evaluating runtime performance across different precision levels.


\section{Profiling Methodology}\label{methodology}
\begin{figure}[t]
    \centering
    \includegraphics[width=\columnwidth]{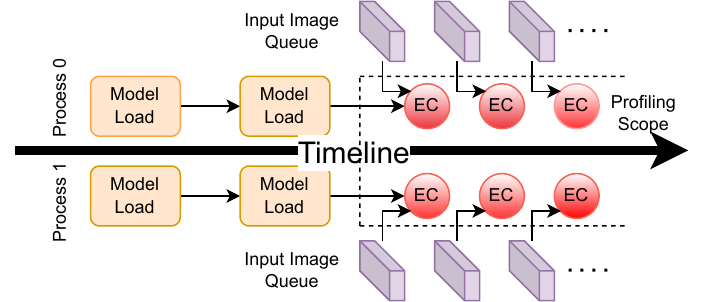}
    \caption{Inference timeline and profiling scope. $EC_i$ denotes execution of $I_i$th image.}
    \label{profiling-scope}
\end{figure}
\noindent
Inference workloads typically involve executing thousands of rapid, homogeneous iterations. Analysing individual iterations in such scenarios poses significant challenges, particularly in multi-process environments where concurrent operations can obscure intricate interactions and dependencies.

\noindent
At first, our objective is to capture raw data on throughput and power consumption with minimal intrusion from the profiler in the runtime. To gain comprehensive insights into system performance and identify bottlenecks, deeper-level data acquisition is necessary, albeit with increased profiling intrusion. Our benchmarking methodology is divided into two distinct phases. In the first phase, we leverage the lightweight \textit{Jetson-Stats} module alongside the parallel execution of $trtexec$ to ensure that the actual inference loop remains unaffected by profiling overhead. Key metrics such as throughput, power consumption, and GPU memory utilisation are collected in this phase. Throughput is collected from \textit{trtexec} tool. Power consumption and GPU memory utilisation are collected from the Jetson-stats tool. The minimal impact of the profiling tools ensures a realistic measurement of system-level performance under sustained workloads. 

\noindent
The second phase employs \textit{NVIDIA Nsight Systems}, a medium-overhead profiler that provides detailed insights into hardware-level performance. The overhead introduced is proportional to the number of metrics collected, allowing a trade-off between granularity and system impact. The second phase focuses on kernel-level metrics, including tensor core utilisation and CUDA execution efficiency, offering a detailed view of GPU operation and CPU-GPU interactions during inference. In the second phase, using the \textit{Nsight Systems} profiler, we introduced an intrusion that reduced throughput by $50\%$. The details of all the collected metrics are discussed in section \ref {collected_metrics}.

\noindent
As depicted in Fig-\ref{profiling-scope}, both phases are preceded by warm-up iterations to stabilise system performance and mitigate transient effects. Profiling runs are conducted over extended durations to comprehensively capture interaction patterns. Notably, \textit{trtexec} pre-enqueues one batch into the input queue, thereby eliminating GPU idling caused by CPU-side image preprocessing. While real-world workloads may experience inter-batch latency, our methodology approximates an upper bound for model throughput under ideal conditions.  

\noindent
This dual-phase approach provides a holistic view of system performance, bridging application-level metrics and low-level hardware insights.

\section{Collected Metrics}\label{collected_metrics}
\noindent
In this section, we describe the metrics we collect during our experiments, 
\begin{table}[htbp]
\label{metrics_info}
\centering
\begin{tabularx}{\columnwidth}{>{\raggedright\arraybackslash}p{1in} >{\raggedright\arraybackslash}X} \midrule
\multicolumn{2}{c}{\textbf{SoC Level Metrics}} \\
\rowcolor{gray!20} \textbf{Metric} & \textbf{Description} \\
Throughput & Total number of images processed in unit time \\
Power & Power consumption in Watt \\ \midrule
\multicolumn{2}{c}{\textbf{GPU Level Metrics}} \\
\rowcolor{gray!20} \textbf{Metric} & \textbf{Description} \\ 
GPU Utilisation & GPU compute time/ total wall time \\
GPU Memory & GPU Memory usage (\%) \\
SM Issue Cycles & SM cycles issued with an instruction issued(\%) \\
SM Active Cycles & SM Cycle with at least 1 warp(\%) \\
TC Utilization & TC active cycles/ Total Cycles(\%) \\
\midrule
\multicolumn{2}{c}{\textbf{Kernel Level Metrics}} \\ 
\rowcolor{gray!20} \textbf{Metric} & \textbf{Description} \\
Launch Stats & Time GPU spends on kernel launch \\
Sync Time & Time GPU spends on synchronising kernels \\
EC Time & Time to execute an Execution context \\
\bottomrule
\end{tabularx}
\caption{Different levels of collected Metrics}
\end{table}
\subsection{SoC-Level Metrics}
We collect two key edge computing metrics—throughput and power consumption—using the tools \textit{trtexec} and \textit{jetson-stats}. Throughput is defined as the number of images processed per second. Throughput and power consumption are critical factors in assessing the trade-offs in offloading inference processes on edge architectures.

\subsection{GPU-Level Metrics}
SoC-level metrics provide a high-level overview but fall short of providing insights into the development and optimisation of deep learning models. GPU-level metrics—such as GPU utilisation, memory usage, SM utilisation, and tensor core utilisation—offer deeper insights into a model’s runtime performance.
\subsubsection{GPU Utilisation and Memory Usage}
\noindent
GPU utilisation measures the proportion of time a process runs on the GPU. TensorRT (TRT) models are designed to use relatively low memory during runtime \cite{tensorrt}.
TensorRT allocates device memory to store the model weights upon loading. Since the TRT model has almost all the weights, its size approximates the amount of GPU memory the weights require \cite{tensorrt}.
\subsubsection{SM Utilisation and Issue Slot Utilisation}
\noindent
A Streaming Multiprocessor (SM) is a fundamental component of NVIDIA GPUs, consisting of multiple CUDA cores that execute instructions in parallel. When a GPU kernel is launched, threads are grouped into blocks(warps) and distributed across SMs by the GPU scheduler. The aim is to keep SMs fully occupied, though at the start and end of execution, under-utilisation may occur due to insufficient thread blocks. \textit{SM active utilisation} is defined as the ratio of SM cycles (where at least one warp is active) to the total elapsed GPU cycles. It reflects how effectively the GPU scheduler parallelises threads.

\noindent
However, active SMs do not always issue instructions; the warp scheduler may stall due to data fetching or other delays. A cycle with no issued instructions is termed a \textit{stalled cycle}, while an \textit{issue cycle} is one in which instructions are issued. \textit{SM issue slot utilisation} is the ratio of issue cycles to elapsed cycles and serves as a lower bound on SM active utilisation.
\subsubsection{Tensor Core Utilisation}
\noindent
Tensor Cores (TC) \cite{tensorcore} are specialised units designed for accelerating matrix multiplication. \textit{TC utilisation} is the ratio of TC cycles to the total number of GPU cycles. Ideally, most operations should be executed on Tensor Cores, with SM instructions dominated by TC instructions.

\noindent
These metrics are crucial for gaining insight into the deployed model's runtime execution, which in turn helps in the optimisation of the DL model, especially when GPU utilisation is high, but SM or Tensor core utilisation is low. They help evaluate the GPU's efficiency in scheduling tasks across computing units.
\subsection{Kernel-Level Metrics}
Kernel-level metrics help identify runtime bottlenecks, particularly in kernel launches and \textit{CudaSynchronization} (CS) event statistics. While synchronisation is necessary to manage branching conditions and asynchronous CUDA operations, it can also introduce runtime overhead. We also measure details of the \textit{ExecutionContext}(EC), which is a class of TensorRT SDK that helps to manage the inference of a single batch. It contains all the states associated with a particular inference invocation; thus, we can have multiple contexts associated with a single engine and run them in parallel \cite{tensorrt}. The entire inference timeline of a DL runtime is the serial execution of \textit{EC} and \textit{CS} events.

\section{Workload Analysis}\label{results}
\subsection{Reference Workload}\label{reference_workloads}
We define the reference workload as a single DL process running inferences with a batch size of $1$. The image size of a batch matches the default for each model: $3\times224\times224$ for \textit{ResNet50} and \textit{FCN\_ResNet50}, and $3\times640\times640$ for \textit{YoloV8n}. Each model is compiled at various mixed weight precision levels: \textit{int8}, \textit{fp16}, \textit{tf32}, and \textit{fp32}. While precision reduction typically impacts accuracy \cite{quantization}, this aspect is outside the scope of our study.
\subsubsection{GPU Memory Usage \& Throughput vs Precision}
\noindent
In a single-process scenario, GPU memory usage depends primarily on the loaded model size and twice the batch size, with the model size being the dominant factor. Since \textit{trtexec} pre-enqueues one batch in advance, the runtime involves one batch being processed while another waits in the queue. As a result, the batch size is doubled to calculate memory usage.

\begin{figure}[t]
    \centering
    \includegraphics[width=\columnwidth]{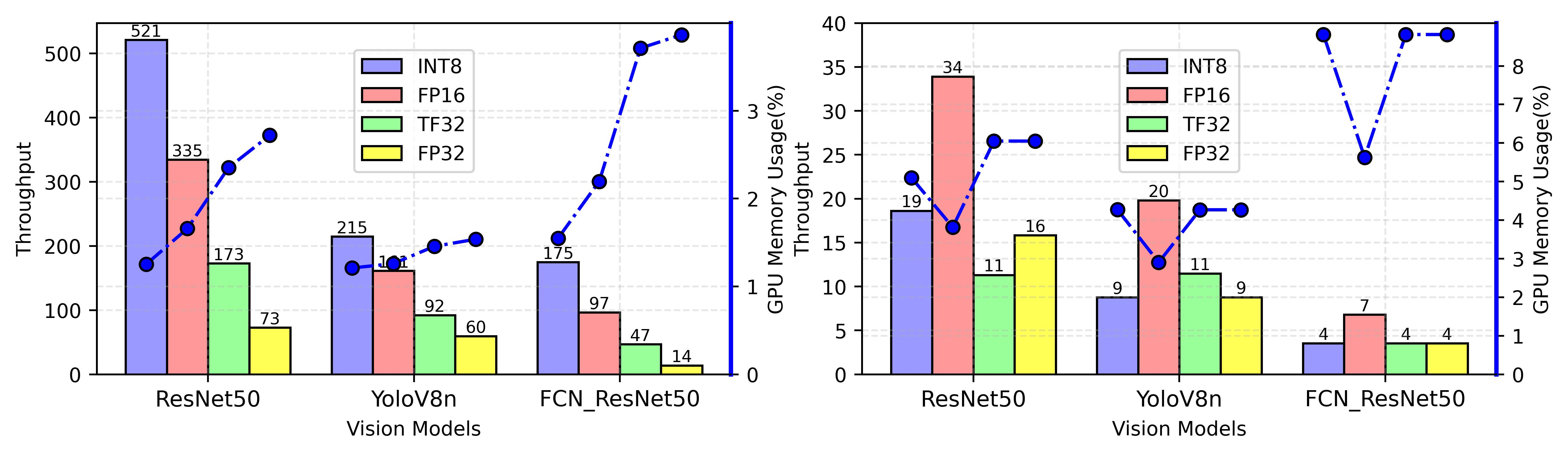}
    \caption{GPU Memory Usage \& Throughput vs. Precision for Vision Workloads: Jetson Orin Nano (Left) and Jetson Nano (Right)}
    \label{throughput_gpu_mem_usage}
\end{figure}

\noindent
As the model precision increases from \textit{int8} to \textit{fp32}, GPU memory usage increases proportionally (Fig.~\ref{throughput_gpu_mem_usage}) on the Jetson Orin Nano. For instance, in \textit{ResNet50} and \textit{FCN\_ResNet50}, the \textit{fp32} models consume $2\times$ more memory compared to their \textit{int8} counterparts, while for \textit{YoloV8n}, the increase is approximately $1.25\times$.
In the case of throughput,\textit{int8} models consistently outperform others. Specifically, \textit{int8} models of \textit{ResNet50} and \textit{FCN\_ResNet50} are $9.75\times$ and $12\times$ faster, respectively, while for \textit{Yolov8n}, the speed-up is around $3\times$.

\noindent
In contrast, the Jetson Nano exhibits significantly lower performance than the Jetson Orin Nano, primarily due to its older GPU architecture and lack of tensor cores, resulting in a substantial performance gap. Notably, on the Jetson Nano, \textit{fp16} models exhibit higher throughput and lower memory consumption than their \textit{int8} counterparts. For example, in the case of $YOLOv8n$, the \textit{fp16} model achieves a throughput of $20$, double that of other precision models, while consuming approximately $50\%$ less GPU memory. The Jetson Nano lacks comprehensive support for \textit{int8} and \textit{tf32} precision across all layers of deep learning models. During TensorRT model creation, unsupported layers default to \textit{tf32} precision, whereas compatible layers are converted to either \textit{int8} or \textit{tf32}. Consequently, because the majority of layers operate in \textit{fp32}, throughput decreases and memory utilisation increases.
\begin{conclusionbox}
    \textit{Deploying $int8$ precision models are beneficial on Jetson orin nano whereas, $fp16$ models are optimal for Jetson nano devices.}
\end{conclusionbox}
\subsubsection{Power Consumption vs Precision}
\begin{figure}[b]
    \centering
    \includegraphics[width=\columnwidth]{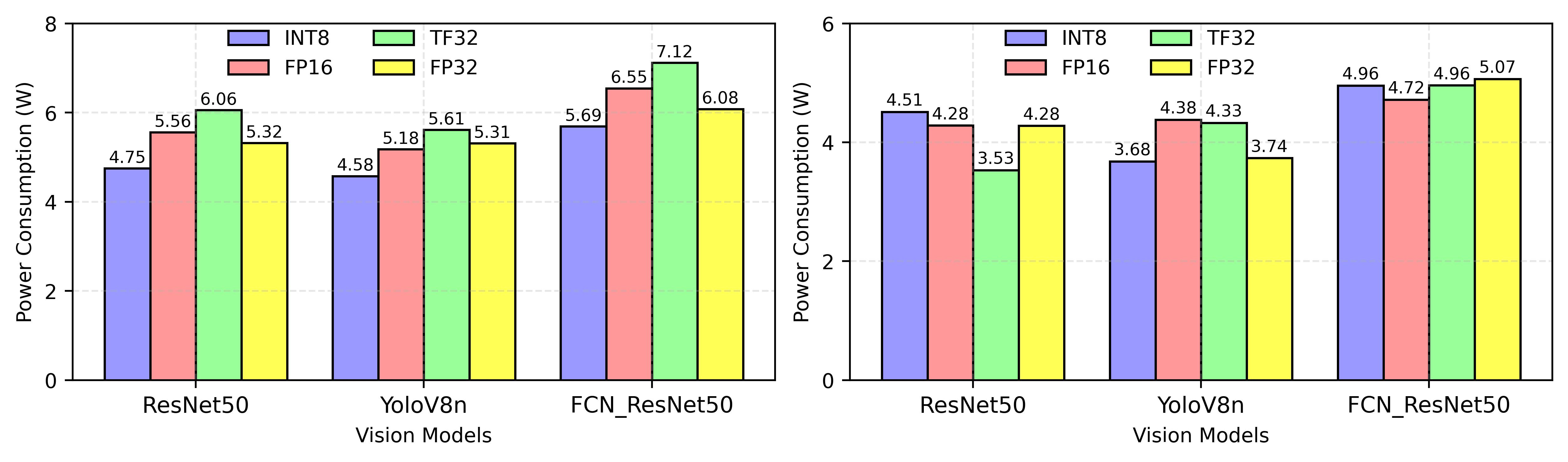}
    \caption{Power Consumption vs Precision for Vision Workloads: Jetson Orin Nano(Left) and Jetson Nano(Right)}
    \label{power_vs_precision_single_batch_single_stream}
\end{figure}

\noindent
For the Jetson Orin Nano, the relationship between precision levels and power consumption is generally proportional, except for \textit{fp32} on the Jetson Orin Nano and \textit{fp16} on the Jetson Nano, as shown in Fig.~\ref{power_vs_precision_single_batch_single_stream}. The \textit{FCN\_ResNet50} model, in particular, shows higher power consumption than the other models. Power consumption increases with precision but notably drops for \textit{fp32} across all models. Interestingly, \textit{fp32} precision models sometimes consume less power than \textit{tf32} or even \textit{fp16} precision models for Jetson Orin Nano. This counterintuitive trend is linked to reduced throughput and the Dynamic Voltage and Frequency Scaling (DVFS) mechanism in the SoC. DVFS is a control mechanism that reduces system frequency when heavy computations risk exceeding thermal and power limits due to high current draw.
\noindent
However, for Orin Nano, power consumption per image also rises with precision for all models, even for $tf32$ precision. For instance, in \textit{FCN\_ResNet50}, the throughput for \textit{fp16}, \textit{tf32}, and \textit{fp32} are $97,\ 47,\ 14$ approximately, respectively, with power consumption at $6.6W,\ 7.1W,\ 6.1W$ approximately. This results in per-image power consumption of $68mW,\ 151mW,\ 435mW$ approximately for \textit{fp16}, \textit{tf32}, and \textit{fp32}, respectively.

\noindent
On the Jetson Nano, power consumption for \textit{int8}, \textit{fp32}, and \textit{tf32} models is similar because most layers process data internally in \textit{fp32}. In contrast, \textit{fp16} models may show slightly higher overall power usage but are significantly more efficient, using the least power per image processed. For example, with the $ResNet50$ model, the power consumption per image is approximately $0.23$ W for \textit{int8}, $0.125$ W for \textit{fp16}, and $0.32$ W for \textit{tf32} precision. This indicates that \textit{fp16} models consume about half the power per image compared to \textit{tf32}.
\begin{conclusionbox}
    \textit{GPU memory usage typically increases when higher precision levels are used. Supported precision formats are more efficient and consume less power per image compared to unsupported formats. In unsupported models, the weights default to \textit{fp32} precision, which results in higher power consumption.}
\end{conclusionbox}
\begin{figure*}[h!]
    \centering
        \includegraphics[width=\textwidth]{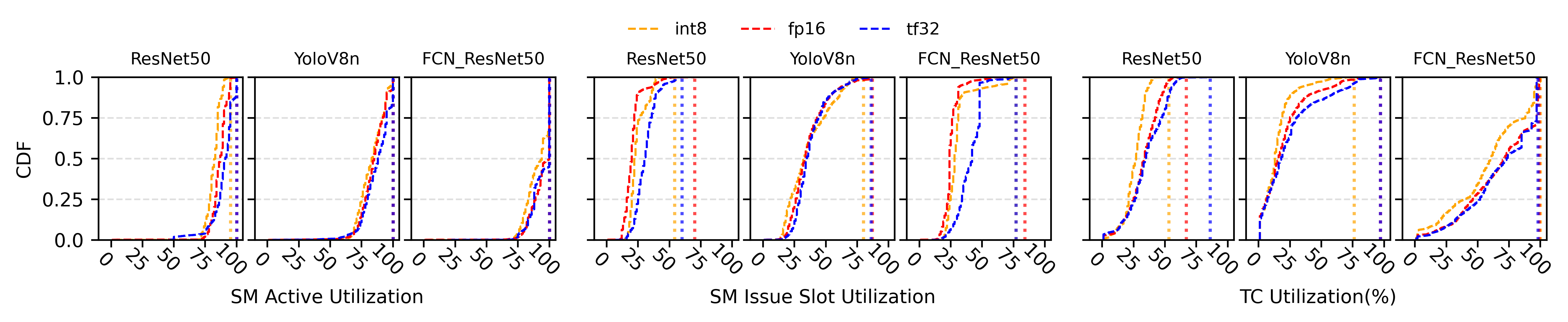}
    \caption{SM Active and Issue Slot \& TC Utilization vs Precision}
    \label{single_batch_single_stream}
\end{figure*}
\subsubsection{SM Utilization \& Issue Slot Utilization}
\noindent
Monitoring Streaming Multiprocessor (SM) utilisation on the Jetson Nano is not feasible with current state-of-the-art profiling tools, as \textit{Nsight Systems} does not support this capability on the device. Consequently, this section exclusively presents results for the Jetson Orin Nano.

\noindent
As shown in Fig-\ref{single_batch_single_stream}, all deep learning models achieve $100\%$ SM active utilisation at some point, with \textit{int8} precision models consistently exhibiting the lowest SM utilisation. For \textit{ResNet50}, SM active utilisation typically ranges between $75$ to $90\%$ across all precisions, with the \textit{tf32} precision rapidly reaching $100\%$ utilisation for $15\%$ of the runtime. In the case of \textit{YoloV8n}, utilisation curves are broader, typically spanning $ 75\%$ to $100\%$. Similar to \textit{ResNet50}, \textit{YoloV8n} in \textit{tf32} precision also reaches $100\%$ utilization for approximately $20\%$ of the runtime. For \textit{FCN\_ResNet50}, the cumulative plot shows SM utilisation predominantly between $75$ to $100\%$ across all precisions, with around $40\%$ of the points reaching $100\%$.

\noindent
However, SM issue slot utilisation remains below $80\%$ for all models, even in \textit{tf32} precision. For \textit{ResNet50}, values concentrate in the $25$ to $40\%$ range, while \textit{YoloV8n} shows a more gradual distribution from $25$ to $75\%$. \textit{FCN\_ResNet50} centres around $25\%$, indicating significant issue stalls despite extended SM activity. This low issue slot utilisation is reflected in throughput, with \textit{ResNet50} achieving the highest and \textit{FCN\_ResNet50} consistently the lowest across all precisions.
\begin{conclusionbox}
    \textit{While all models reach high SM utilisation, the low SM issue slot utilisation, particularly in \textit{FCN\_ResNet50}, highlights significant instruction stalls, directly impacting throughput }
\end{conclusionbox}

\subsubsection{Tensor Core Utilization}
\noindent
Jetson Nano lacks support for Tensor Cores(TCs). That's why, in this section, we also only present the results for Jetson Orin Nano.

\noindent
As shown in Fig.~\ref {single_batch_single_stream}, \textit{ResNet50} and \textit{YoloV8n} models exhibit steep CDFs, indicating rapid attainment of high CDF values. However, TC utilisation never reaches $100\%$ at any precision level. For \textit{ResNet50} at \textit{int8} precision, utilisation remains below $50\%$, mostly around $25\%$. \textit{YoloV8n's} utilisation is concentrated below $20\%$, suggesting it uses TCs less frequently than \textit{ResNet50}, despite potentially higher peak utilisation.

\noindent
In contrast, \textit{FCN\_ResNet50} has a more gradual CDF curve, especially for \textit{int8}, approximating a straight line. CDFs for \textit{fp16} and \textit{tf32} show near-vertical slopes, with $40\%$ of values approaching $100\%$, indicating extensive TC use at these precisions.

\noindent
Interestingly, higher TC utilisation does not always equate to higher throughput. For instance, \textit{FCN\_ResNet50}, despite high TC utilisation, does not exhibit superior throughput. In contrast, \textit{int8} precision, with lower TC utilisation, achieves higher throughput, suggesting factors such as memory bandwidth may also influence performance.
\begin{conclusionbox}
    \textit{TC utilisation for \textit{int8} models is lower than for \textit{fp16} and \textit{tf32}, indicating that even at high utilisation, throughput may be constrained by memory bandwidth and non-TC instructions.}
\end{conclusionbox}
\begin{figure}[tbp]
\centering
\includegraphics[width=\columnwidth]{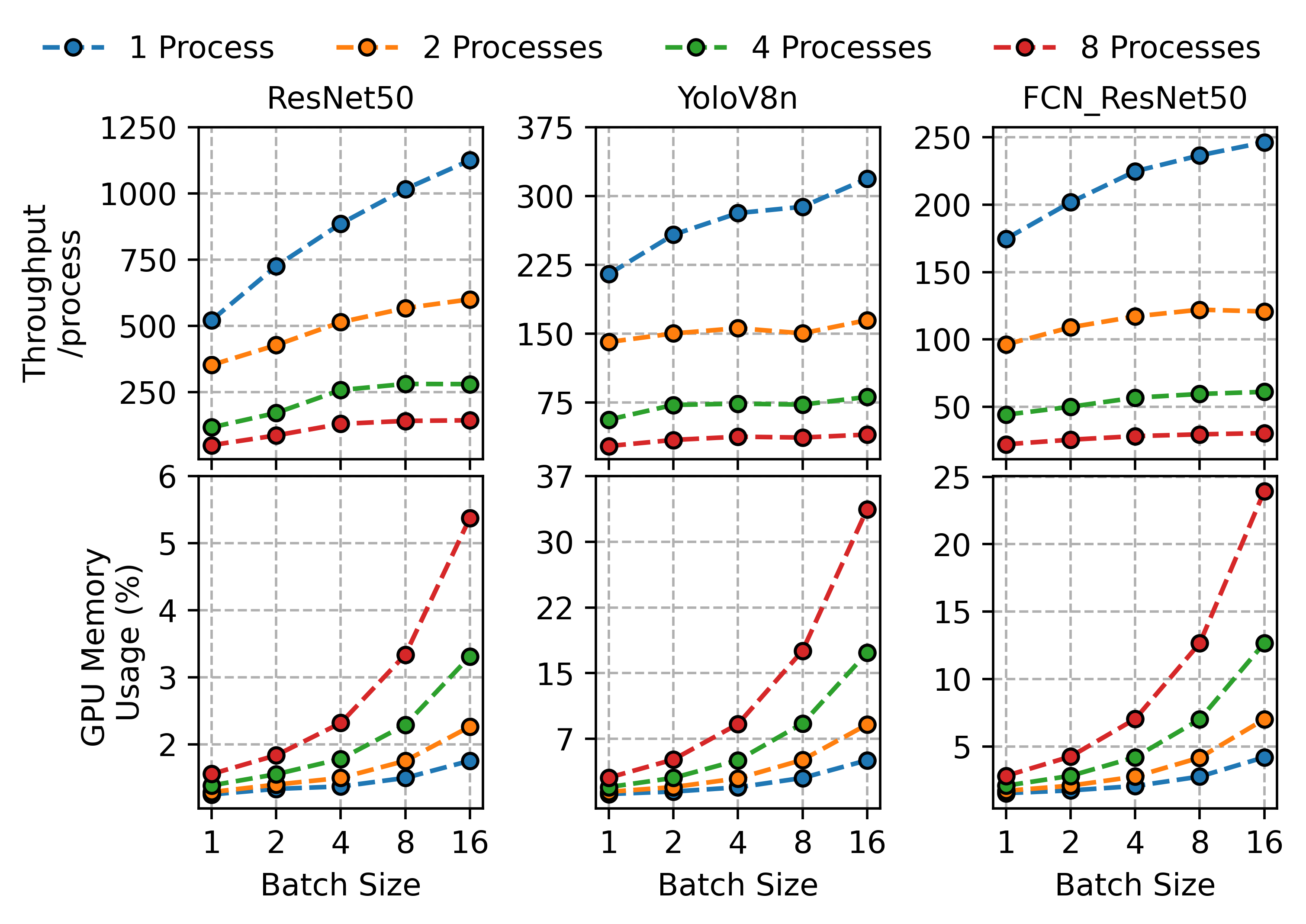}
\caption{GPU Memory Usage (\%) and T/P for \textit{int8},ResNet50, FCN\_ResNet50, and YoloV8n models on Jetson Orin Nano}
\label{gpu_memory_throughput_int8_all_models}
\end{figure}
\subsection{Concurrent Workloads} \label{concurrent_workloads}
The previous section analysed the impact of different precision formats on various performance metrics. However, the effect of concurrent processes on edge GPU systems remains unclear. Our observations show that while the Jetson Nano struggles with throughput and memory usage, the Jetson Orin Nano performs better in these areas. For the Jetson Nano, using $fp16$ precision provides the best throughput, while the $int8$ precision delivers optimal performance on the Jetson Orin Nano. Given the limited capabilities of these GPUs, all subsequent experiments and results will focus on varying the number of concurrent processes. The specific number of concurrent processes will depend on the hardware configuration, the deployed deep learning model, and the available GPU memory.
\setcounter{subsubsection}{0} 
\subsubsection{GPU Memory Usage \& Throughput}
\noindent
In this section, we will discuss the throughput per process (T/P) and GPU memory usage for concurrent workloads. Instead of using the overall system throughput, we consider the T/P metric, as it provides a clear overview of the throughput we expect for each process, which is the main concern in the edge paradigm. In a multi-process scenario, GPU memory usage can be calculated as the number of processes multiplied by the sum of the model size and twice the batch size. This is because each process independently allocates its own memory.
\begin{figure}[tbp]
\centering
\includegraphics[width=\columnwidth]{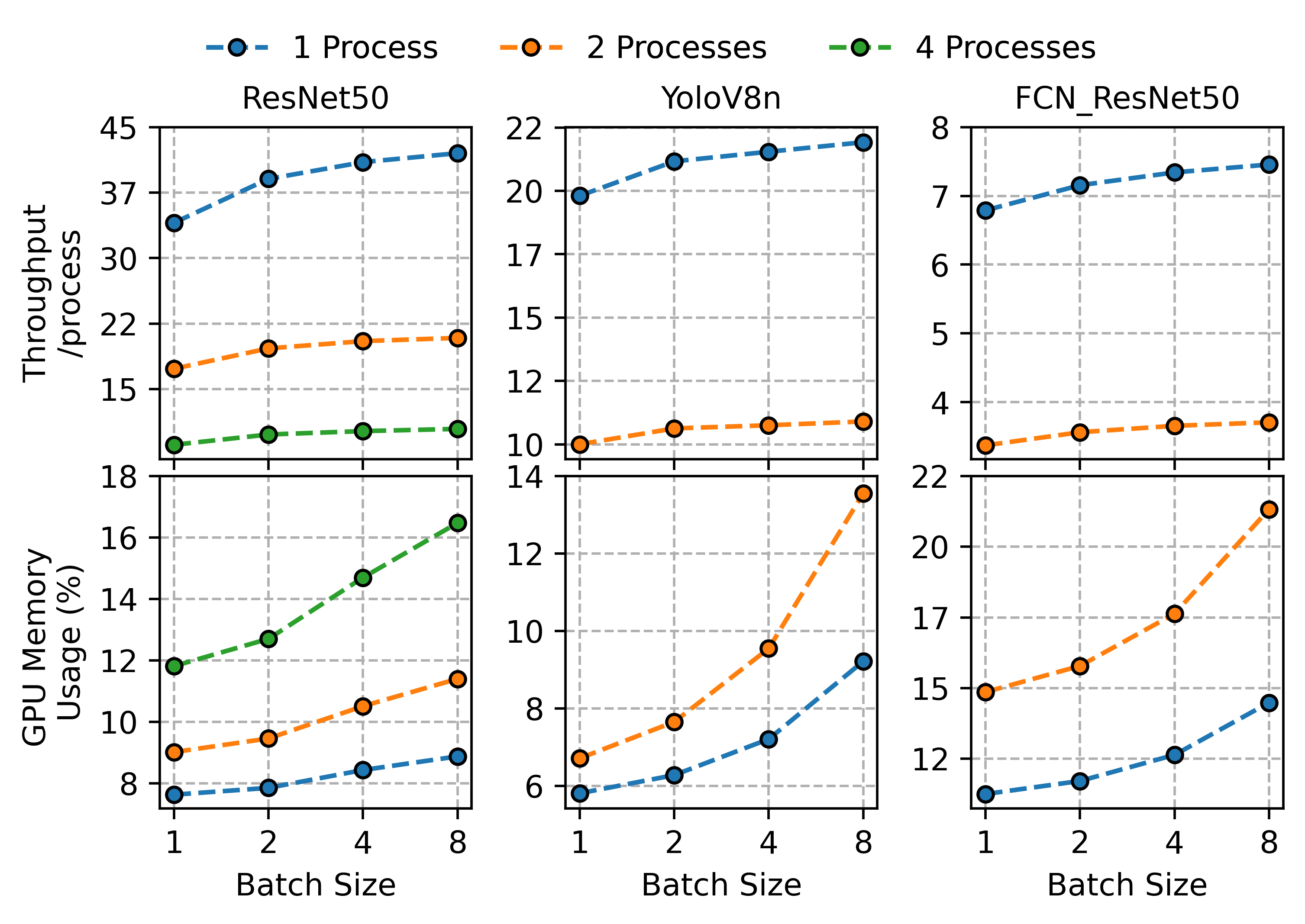}
\caption{GPU Memory Usage (\%) and T/P for \textit{fp16} ResNet50, FCN\_ResNet50, and YoloV8n models on Jetson Nano}
\label{gpu_memory_throughput_int8_all_models_1}
\end{figure}

\noindent
In Fig.~\ref{gpu_memory_throughput_int8_all_models} and Fig.~\ref{gpu_memory_throughput_int8_all_models_1}, it is evident that T/P increases with increasing batch size. However, T/P declines as the number of concurrent processes increases. For instance, consider the \textit{YoloV8n} model: the T/P metric for a single batch in a single-process scenario was around $210$, increasing to $320$ for a batch size of $16$ on the Jetson Orin Nano. For Jetson Nano, the \textit{YoloV8n} model, the T/P metric for a single-batch and single-process scenario was $20$, which increased to $22$ for an $8$ batch size. Interestingly, the increase in throughput/process is not proportional to the increase in batch size. But when we increase the number of processes from $1$ to $8$, the T/P metric decreases to nearly $10$. Similar kinds of behaviour can be seen across all the models.

\noindent
It is important to note that we cannot increase the number of concurrent processes to $8$ or more. This factor is very much dependent on the unified RAM available on the device. For example, for the \textit{ResNet50} model, we can safely deploy up to $4$ processes on the Jetson Nano, whereas for the \textit{FCN\_ResNet50} model, we cannot deploy $4$ processes without causing memory shortages that lead to system reboots.

\noindent
Meanwhile, GPU memory usage is proportional to both batch size and the number of concurrent processes. Notably, the increase is sharp when we increase the process count from $1$ to $8$. For example, the \textit{YoloV8n} model uses less than $10\%$ of GPU memory for a single process with a batch size of $8$, whereas it requires more than $35\%$ of GPU memory when running 16 processes concurrently, indicating a sharp $3.5\times$ increase in GPU memory usage.  

\subsubsection{Power Consumption}
\begin{figure}[t]
    \centering
    \includegraphics[width=\columnwidth]{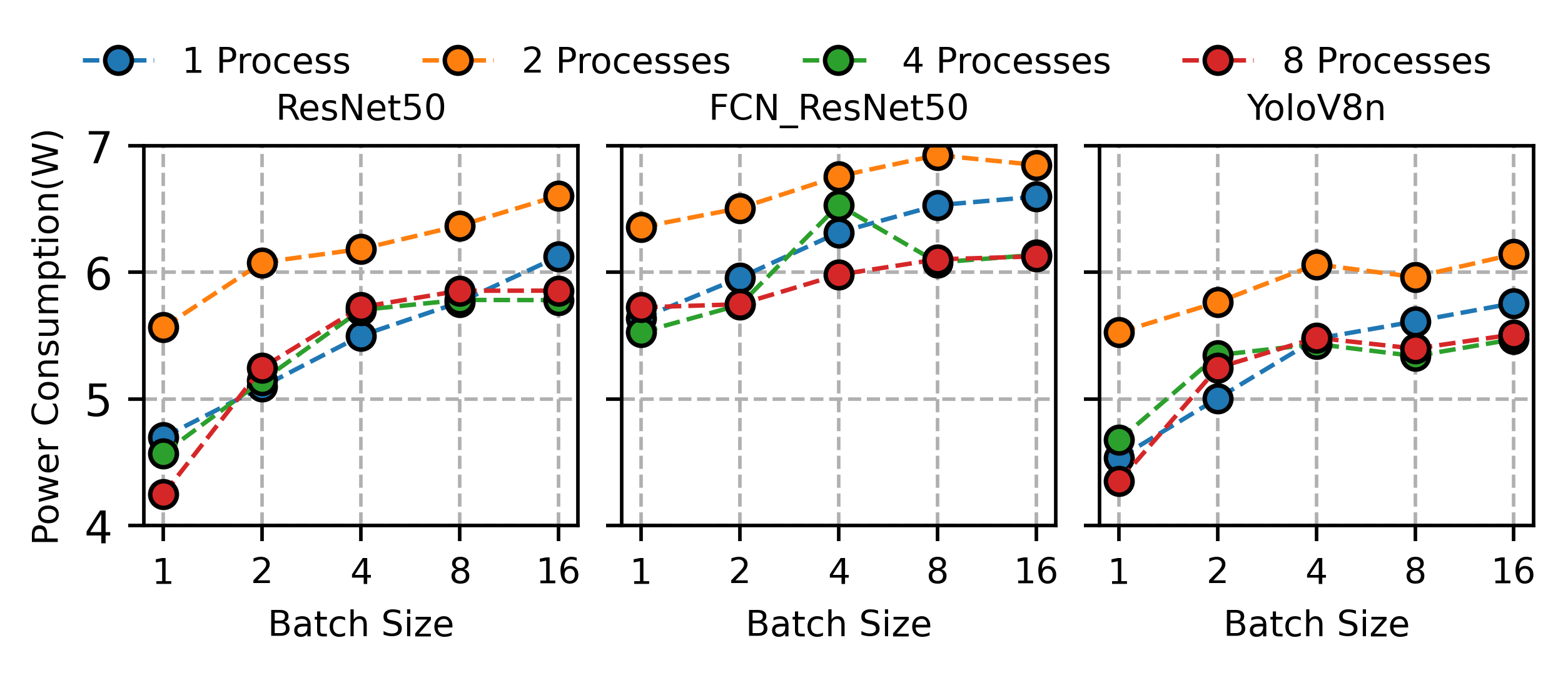}
    \caption{Power Consumption for \textit{int8} \textit{ResNet50}, \textit{FCN\_ResNet50} and \textit{Yolov8n} model on Jetson Orin Nano}
    \label{gpu_energy_consumption_all_model}
\end{figure}

\noindent
In Fig.~\ref {gpu_energy_consumption_all_model}, the power consumption appears arbitrary across different batch sizes and numbers of processes. At first glance, it appears that power consumption increases with batch size. For example, for Orin Nano, across all the models, the power consumption in the 1-process scenario increased from batch size $1$ (approximately $4.75W,\ 5.51W,\ 4.51W$) up to batch size $16$ (approximately $6.17W,\ 6.71W,\ 5.89W$) for \textit{ResNet50}, \textit{FCN\_ResNet50}, and \textit{YoloV8n} models, respectively). However, upon closer inspection of Fig.~\ref{gpu_energy_consumption_all_model}, it is evident that although energy consumption increases with the number of processes, the increments are not smooth and sometimes deviate from the expected behaviour.

\begin{figure}[b]
    \centering
    \includegraphics[width=\columnwidth]{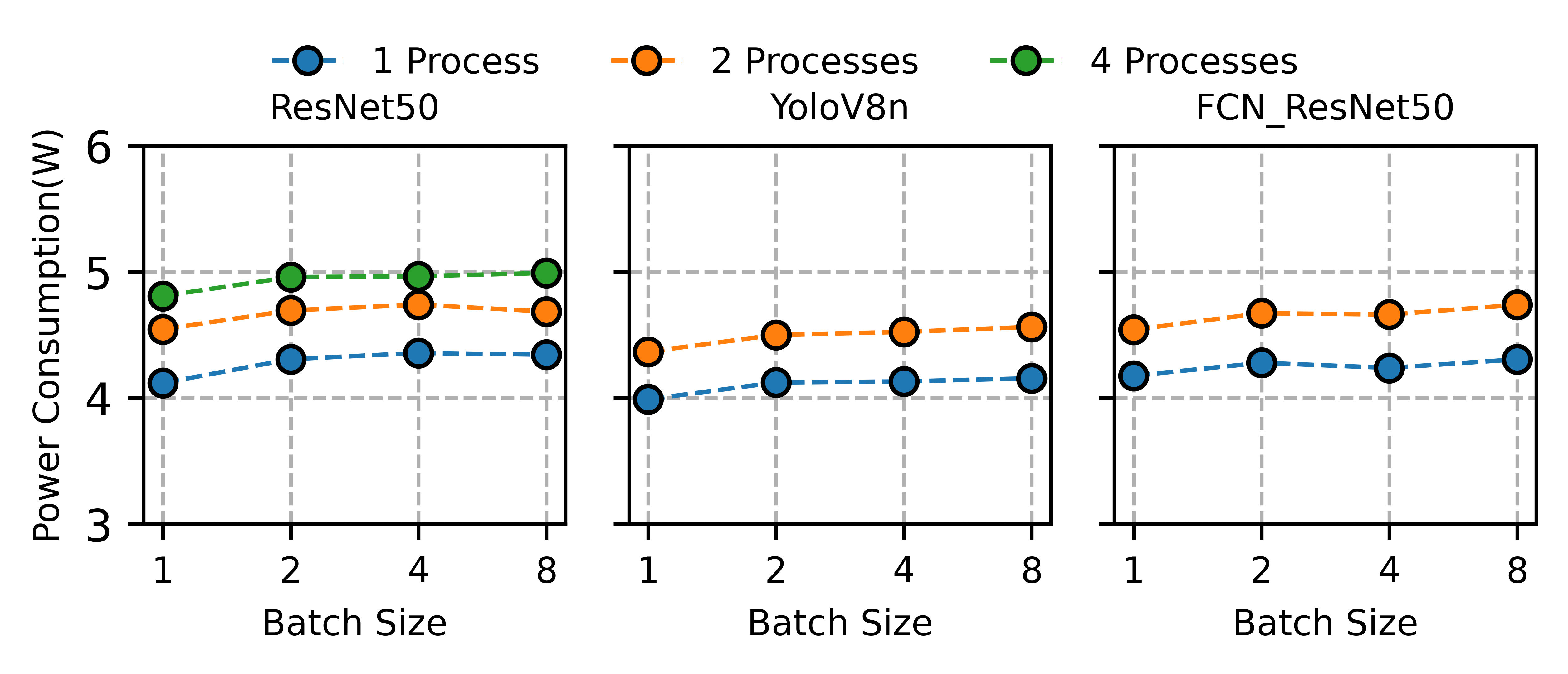}
    \caption{Power Consumption for \textit{fp16} \textit{ResNet50}, \textit{FCN\_ResNet50} and \textit{Yolov8n} model on Jetson Nano}
    \label{gpu_energy_consumption_all_model_jetson_nano}
\end{figure}
\noindent
In some executions, power consumption is higher with fewer processes than with more. For instance, for \textit{FCN\_ResNet50}, the power consumption in the $2$-process scenario is slightly higher across all batch sizes than in the $4$- or $8$-process scenarios. A similar pattern can be observed for the \textit{YoloV8n} model. Additionally, it is notable that for any batch size in any number of processes, the \textit{FCN\_ResNet50} model consistently consumes more energy than its two counterparts.
In Fig.~\ref{gpu_energy_consumption_all_model_jetson_nano}, the Jetson Nano's power consumption is intuitive and follows a clear trend. Power consumption increases with increasing batch size and number of processes, which is somewhat similar to the Jetson Orin Nano. For example, the $FCN\_ResNet50$ model consumes $4.18$, $4.28$, $4.23$, and $4.31$ W power for $1,2,4,8$ batch size for $1$ process configuration. 

\noindent
From the two plots above, it is clear that power consumption never exceeds a certain value, $7W$ for Jetson Orin Nano and $5W$ for Jetson Nano, and that it is somewhat proportional to throughput. But when power consumption exceeds a certain threshold, the device reduces throughput to keep power in check. This method is DVFS and introduces nonlinearity, as shown in Fig.~\ref{gpu_energy_consumption_all_model}.
\begin{conclusionbox}
    \textit{As batch sizes increase, both T/P and GPU memory usage rise. However, T/P declines as the number of concurrent processes increases, while GPU memory usage continues to grow. Power consumption shows a non-linear pattern: it initially rises with batch size and process count, but fluctuates.}
\end{conclusionbox}
\subsubsection{SM Utilization \& Issue Slot Utilization}
\begin{figure*}[htbp]
    \centering
        \includegraphics[width=\textwidth]{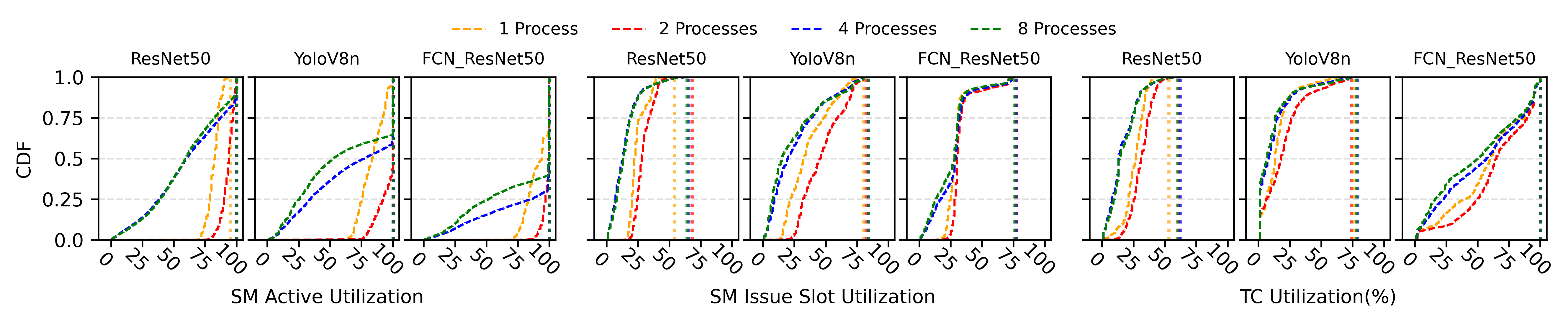}
    \caption{SM Active and Issue Slot \& TC Utilization vs Concurrent Processes}
    \label{single_batch_multi_stream}
\end{figure*}

\noindent
In Fig.~\ref{single_batch_multi_stream}, we present a comparative analysis of single-batch, multi-process scenarios, focusing on SM active and issue-slot utilisation. The cumulative plots for SM utilisation reveal steep curves for $1$ and $2$ concurrent processes, particularly with the \textit{ResNet50} model, where utilisation is predominantly in the $80-100\%$ range. For $4$ and $8$ processes, the curves show a more gradual slope, a trend consistent across all models. Notably, the $2$-process configuration often reaches $100\%$ SM utilisation, especially with \textit{FCN\_ResNet50}, where over $60\%$ of runtime hits this peak.

\noindent
For issue slot utilisation, $1$- and $2$-process configurations exhibit higher levels than $4$- or $8$-process configurations. The \textit{ResNet50} and \textit{FCN\_ResNet50} models cluster around $25\%$ utilisation, while \textit{YoloV8n} shows a broader distribution. No model exceeds $80\%$ issue-slot utilisation, with the average across all scenarios at around $25\%$.
Due to space constraints, the effects of increasing batch sizes have been omitted. However, it is important to note that increasing the batch size slightly increases SM and issue slot utilisation.
\begin{conclusionbox}
    \textit{SM active utilisation increases with increasing number of processes, often reaching $100\%$ for larger numbers of processes, but issue slot utilisation remains stable at around $25\%$ on average across the models}.
\end{conclusionbox}
\subsubsection{Tensor Core Utilisation}
\noindent
In Fig-\ref{single_batch_multi_stream}, the Tensor Core (TC) utilisation for the \textit{ResNet50} model consistently hovers around $25\%$ on average when the process count is $1$ or $2$. However, this utilisation declines to approximately $15-20\%$ as the process count increases to 4 and 8. A similar pattern is observed for the \textit{YoloV8n} model, with an average utilisation of around $30\%$. Notably, the TC utilization for \textit{ResNet50} and \textit{YoloV8n} never exceeds $50\%$ and $75\%$, respectively. Interestingly, the maximum utilisation value tends to increase with the number of processes.
For the \textit{FCN\_ResNet50} model, the cumulative plot exhibits a more gradual incline, with a noticeable clustering near the $100\%$ utilisation mark. However, it is important to note that higher TC utilisation does not necessarily correlate with increased throughput. This disparity is likely due to TC issue stalls and low TC issue slot utilisation. Furthermore, it can be inferred that these stalls become more pronounced as the number of processes increases.
\begin{conclusionbox}
    \textit{TC utilisation increases with higher process counts, but for most cases it stays nearly $30\%$}
\end{conclusionbox}

\section{Kernel Level Performance Analysis}\label{kernel_level}
\noindent
To better understand the findings presented in Section-\ref{results}, examining the complete inference runtime timeline is essential. In this section, we focus exclusively on analysing the performance of the $ResNet50$ model with $int8$ precision on the Jetson Orin Nano and the $fp16$ precision model on the Jetson Nano. Insights for other models and NVIDIA-based devices can be readily derived using the analysis below and information from other sections.

\begin{figure}[t]
    \centering
    \includegraphics[width=\columnwidth]{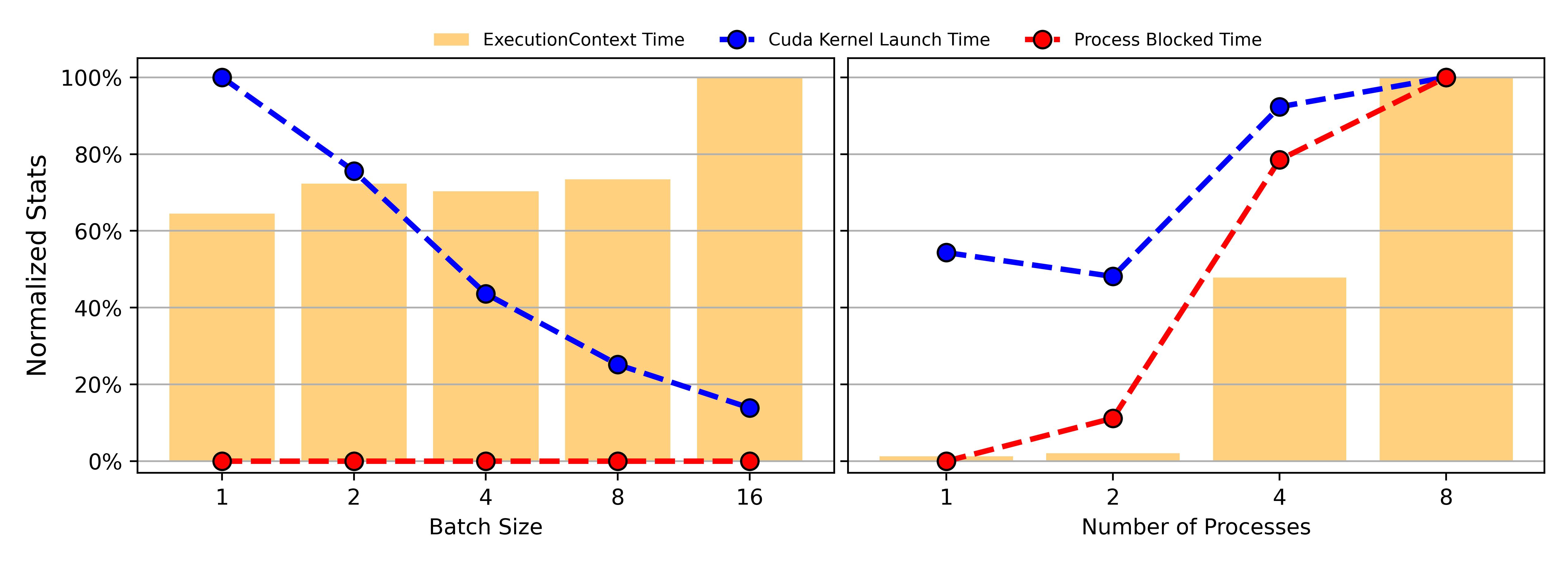}
    \caption{Comparison of GPU and CPU Events for \textit{ResNet50} $int8$ on Jetson Orin Nano: vs. Batch Sizes (Left), vs. Process Counts (Right)}
    \label{resnet_execution_context}
    \vspace{-5mm}
\end{figure}

\noindent
This timeline comprises a sequence of TensorRT \textit{ExecutionContext} (EC) events interspersed with \textit{Cuda Synchronisation} (CS) events. The timeline can be expressed as $\sum_{i}EC_i+\sum_{j}CS_j$, where $EC_i$ represents the average duration of an execution context and $CS_j$ denotes the average duration of a CUDA synchronisation event. As illustrated in Fig-\ref{resnet_execution_context}, for configurations with $1$ or $2$ concurrent processes, the $EC_i$ duration remains stable and relatively short (\textit{1–2 ms}). However, with $4$ or $8$ concurrent processes, the $EC_i$ duration increases significantly, by approximately $30\times$ and $70\times$, respectively, which seems exponential to the number of processes.

\noindent
Interestingly, increasing the batch size has a minimal effect on the $EC_i$ duration, with only a slight increase observed at a batch size of 16. An EC with a batch size of $x$ signifies the simultaneous inference of $x$ images, thereby enabling higher throughput with only a marginal rise in processing time. This phenomenon explains why increasing the batch size improves throughput, albeit with diminishing returns, whereas increasing the number of concurrent processes reduces throughput. Notably, variations in batch size, particularly increases, have a limited impact on scheduling; in some cases, a larger batch size may even enhance GPU scheduling efficiency.

\noindent
A more detailed analysis reveals that each execution context ($EC_i$) is influenced by a multitude of events occurring at both the kernel and CPU levels. Each $EC_i$ can be further represented as a sequence of GPU kernel launches, CPU thread initiations, and similar operations. Formally, this can be expressed as $EC_i = \sum_{l}(K_l + T_l + C_l + B_l)$, where $K_l$ represents the time required to launch a GPU kernel, $T_l$ denotes the time needed to initiate a CPU thread following preemption, $C_l$ signifies the actual computation time and $B_l$ the average blocking time. Blocking time represents the time the process gets blocked. This type of timeline happens due to the \textit{big.LITTLE} \cite{arm_big_little} architecture employed by Arm CPUs \cite{arm_ref_manual}. In this architecture, we have two clusters of CPU cores. One of the clusters handles the heavy load, and the other usually handles the light load. For Jetson Orin Nano, 3 CPU cores are dedicated to heavy loads(in our case, it is the inference workloads), whereas for Jetson Nano, this number is 2.
\noindent
For the Jetson Orin Nano platform, a notable phenomenon occurs when the number of concurrent processes reaches four or more. The processes begin operating in a time-sharing mode, resulting in preemption and deferred scheduling. This behaviour can also lead to the preemption of specific execution contexts ($EC_i$). Based on this, several key observations can be made:
\begin{enumerate}  
    \item The cumulative process blocking time,$\sum_{l}b_l$ rises significantly with an increasing number of processes. For scenarios with one or two processes, the blocking time remains negligible. However, in configurations with four or eight processes, the blocking time becomes a dominant factor influencing $EC_i$, with individual blocking intervals ($b_l$) typically ranging from $1-2\text{ms}$.  

    \item The total time required for CPU thread rescheduling ($\sum_{l}T_l$) and GPU kernel launches ($\sum_{l}K_l$) increases as the number of processes grows. While individual GPU kernel launches are typically brief, taking approximately $K_l=20-100 \mu s$, the cumulative overhead becomes substantial in multi-process scenarios due to frequent context switching.  

    \item For process counts of four or more, the frequent migration of processes across CPU cores leads to a significant rise in cache miss rates, particularly at the $L1$ and $L2$ levels. This increase stems from the loss of temporal and spatial data locality caused by process migration. Consequently, the total computation time ($\sum_{l}C_l$) also increases, further extending the execution context duration.  
\end{enumerate}  

\begin{figure}[t]
    \centering
    \includegraphics[width=\columnwidth]{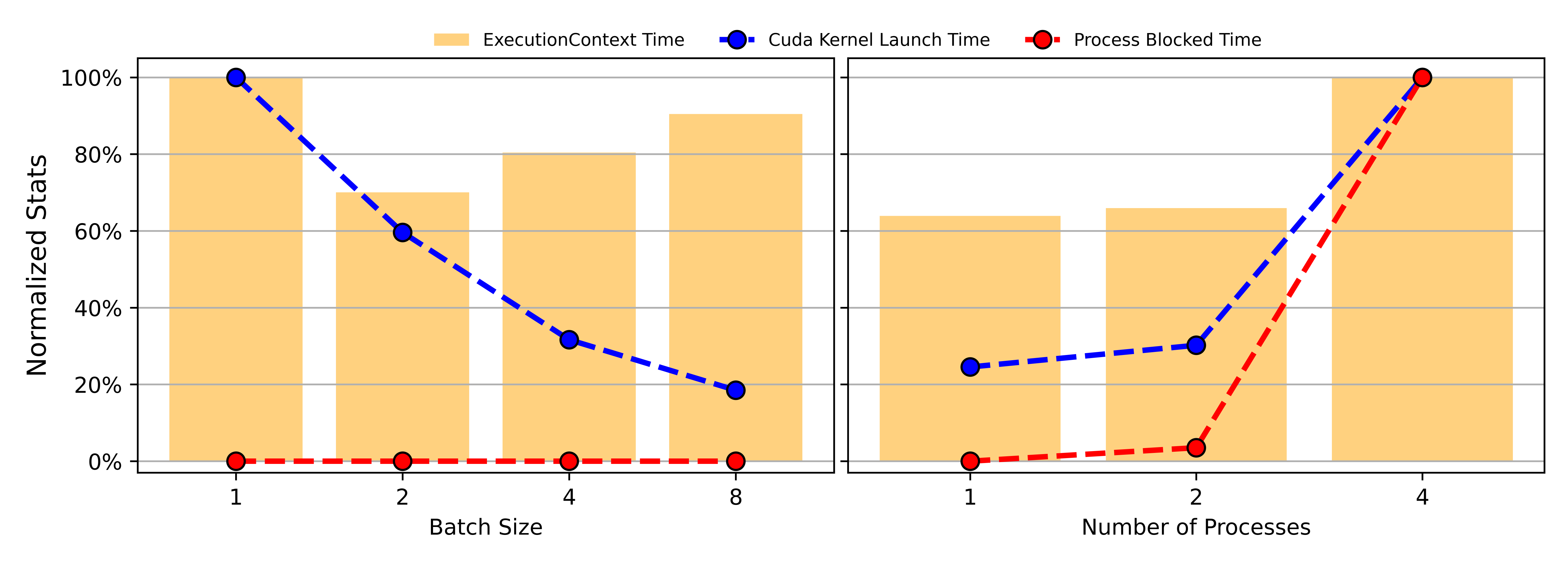}
    \caption{Comparison of CPU and GPU events for \textit{ResNet50} \texttt{fp16} model on Jetson Nano, with varying batch sizes (left) and process counts (right).}
    \label{resnet_execution_context_jetson_nano}
\end{figure}

\noindent
Similar trends are observed in the case of the Jetson Nano (refer to Fig-\ref{resnet_execution_context_jetson_nano}). The execution context duration ($EC_i$) remains largely invariant with respect to batch size, whereas the CUDA kernel launch time decreases noticeably as batch size increases. This pattern persists even as the number of processes increases. When the process count exceeds half of the available CPU cores (e.g., four processes in this case), the execution context duration ($EC_i$) experiences a significant increase, approximately doubling.

\noindent  
These findings highlight that as the number of processes scales, the duration of $EC_i$ increases due to a combination of blocking times, thread and kernel launch delays, and elevated computation times resulting from inefficient cache utilisation. These observations are not restricted to the Jetson Orin Nano platform but are indicative of broader trends in systems where processes contend for shared CPU and GPU resources, particularly in NVIDIA Jetson GPUs.

\begin{conclusionbox}
    \textit{If the number of processes is equal to or fewer than half the available CPU cores, the execution context (EC) duration remains stable, and preemption of the EC does not occur. However, when the number of processes exceeds this threshold, both the EC duration and kernel launch time increase, often exhibiting exponential growth as the process count rises. Notably, employing larger batch sizes helps stabilise the EC duration and reduces the kernel launch time.}
\end{conclusionbox}

\section{Conclusions}\label{conclusions}
\noindent
To achieve optimal performance in inference workloads, it is essential to balance the utilisation of architectural resources, particularly in edge computing scenarios. Performance can be evaluated from multiple perspectives. In the context of the edge computing paradigm, a critical decision must be made: whether to execute tasks at the edge or offload them to the cloud. For instance, in cloud-based environments (such as those utilising NVIDIA A40 GPUs), a single \textit{YoloV8n} model can process approximately \textit{1000+} images per second with $fp16$ precision. However, network delays—including transmission, propagation, and processing—negatively impact the overall throughput experienced by the user.

\noindent
Instead of manual trial and error with QoS requirements(for example, determining optimal number of concurrent processes, optimal number of batch sizes, etc.), we can make decisions based on this type of analysis. Some parts of the computations can either be offloaded to the cloud or distributed horizontally by adding more edge AI accelerators for load balancing. In general, a cloud-edge co-inference system could be designed to accommodate the load without explicitly employing complicated scheduling algorithms. 

\noindent
This analysis also provides valuable insights into the internal workings and resource utilisation of GPU hardware, as well as potential performance bottlenecks. While increasing GPU memory capacity undeniably boosts computational power and throughput, it is important to note that, even with efficient memory utilisation, other internal components—such as SMs, Tensor Cores, and CPU control mechanisms—substantially influence overall performance. Despite achieving $99\%$ GPU utilisation, $100\%$ Tensor Core utilisation, and over $50\%$ SM utilisation, performance may still be constrained by kernel operations and CPU-side optimisations aimed at power management. Our research suggests that, in inference workloads, GPU architecture performance is heavily influenced by the interplay between CPU and GPU scheduling. This underscores the need for further investigation into optimal thread scheduling strategies and the development of programmable architectures that are specifically tailored to accelerate deep learning workloads.

\section*{Acknowledgement}\label{acknowldgement}
\noindent
The research work presented in this article has been supported by the European Commission under the Horizon Europe Programme and the OASEES project (no. 101092702).

\bibliographystyle{unsrt}  
\bibliography{references}

\end{document}